\documentclass[12pt]{article}
\usepackage{epsfig,latexsym,colordvi}
\usepackage{amssymb, amsmath, mathbbol}
\usepackage[centerlast,footnotesize]{caption2}

\newcommand{\lapprox}{\raisebox{-0.5ex}{$\
\stackrel{\textstyle<}{\textstyle\sim}\ $}}

\usepackage{graphicx}
\newcommand{\One}{1\kern-4.5pt1}

\newcommand{\be}{\begin{equation}}
\newcommand{\ee}{\end{equation}}

\def\lesim{${\lower 2pt\hbox{$\scriptstyle
<$}\atop\raise 4pt\hbox{$\scriptstyle\sim$}}$} 
\def\grsim{${\lower2pt\hbox{$\scriptstyle >$} \atop\raise4pt\hbox 
{$\scriptstyle\sim$}}$} 

\begin{document}
\begin{center}
\vskip 5mm
{\LARGE
Monte Carlo simulation of monolayer graphene at non-zero temperature}
\vskip 1.5 cm
{\bf Wesley Armour$^{a,b}$, Simon Hands$^c$, and Costas Strouthos$^d$}
\vskip 0.3 cm
$^a${\em Diamond Light Source, Harwell Campus,\\
Didcot, Oxfordshire OX11 0DE, United Kingdom}
\vskip 0.3 cm
$^b${\em Institute for the Future of Computing, Oxford Martin School, 
\\Oxford e-Research Centre, 7 Keble Road, 
Oxford OX1 3QG, United Kingdom} \\
\vskip 0.3 cm
$^c${\em Department of Physics, College of Science, Swansea University, \\
Singleton Park, Swansea SA2 8PP, United Kingdom} \\
\vskip 0.3 cm
$^d${\em Computation-based Science and Technology Research Center,  \\
The Cyprus Institute, 1645 Nicosia, Cyprus.}
\end{center}

\vskip 0.9 cm
\begin{center}
{\bf Abstract}
\end{center}

\noindent
We present results from lattice simulations of a monolayer graphene model 
at non-zero temperature. 
At low temperatures for sufficiently strong coupling the model
develops an excitonic condensate of particle-hole pairs corresponding
to an insulating phase.
The Berezinskii-Kosterlitz-Thouless phase transition temperature
is associated with the value of the coupling where the critical exponent $\delta$ 
governing the response of the order parameter at criticality to an external
source
has a value close to 15.
The critical coupling on a lattice with temporal extent $N_t=32$ ($T=1/(N_t
a_t)$ where $a_t$ is the temporal lattice 
spacing) and spatial extent $N_s=64$ is very close 
to infinite coupling. 
The value of the transition temperature normalized with the zero temperature
fermion mass gap $\Delta_0$
 is given by $\frac{T_{BKT}}{\Delta_0}=0.055(2)$. 
This value provides an upper bound on the transition temperature, 
because simulations closer to the continuum limit where the full $U(4)$ symmetry
is restored may result in 
an even lower value. 
In addition, we measured the helicity modulus $\Upsilon$ and the fermion
thermal mass $\Delta_T(T)$, the latter providing evidence for a pseudogap phase
with $\Delta_T>0$ extending to arbitrarily high $T$.  Analysis of the
dispersion relation suggests that the Fermi velocity is not sensitive to
thermal effects.

\newpage

\section{Introduction}

The impact of electron-electron interactions on the physics of graphene is an
important focus of current study (for recent reviews see \cite{Neto}).
There are simple
arguments why an ``independent quasiparticle'' picture may not be adequate
for certain properties. Firstly, since the carrier density of states
vanishes in undoped graphene (the zero energy condition is only satisfied at two
isolated ``Dirac points'' in the first Brillouin zone), the effects of screening
are much less in graphene than in a conventional conductor, the main
contribution coming from electron-hole pairs which increase the effective
dielectric constant of the medium in a fashion entirely analogous to vacuum
polarisation in QED. This means that the interaction between charged carriers
remains Coulombic, i.e. long-ranged $\propto r^{-1}$. Secondly, the relative
importance of quantum corrections, parametrised by the fine structure constant
$\alpha$,
is much greater than in conventional QED, because $\alpha_{\rm
eff}={e^2\over{4\pi\varepsilon\hbar v_F}}$, where $v_F\approx{c\over300}$ is the
Fermi velocity and $\varepsilon$ the
dielectric permittivity of the underlying substrate: hence $\alpha_{\rm
eff}=\alpha{c\over v_F}\sim O(1)$, and its
value depends on the substrate, taking a maximum
value 2.16 for suspended graphene.

These considerations have motivated the study of an effective $(2+1)d$ relativistic field
theory with $N_f$ fermion flavours for the low energy electronic excitations
($N_f=2$ for monolayer graphene) and
an instantaneous Coulomb interaction between conserved charges, to
be reviewed in Sec.~\ref{sec:formulation} below~\cite{Khveshchenko1,Gorbar,Son}.
For sufficiently strong coupling the theory describes a quantum
critical point (QCP) at
$T=0$ separating a semimetal phase in which charge carriers remain ungapped,
from an insulating phase in which electron-hole exciton pairs condense in the
ground state inducing a gap at the Dirac points.
It is conceivable that the properties of the QCP dominate the effective
description of low-energy charge transport in graphene irrespective of whether
the semimetal or insulating phase is physically realised.

Since the theory is strongly interacting, various non-perturbative approaches
have been applied, including Monte Carlo simulation of an effective lattice
field theory postulated to belong to the same universality class at the QCP.
In a series of papers,
Drut and L\"ahde~\cite{Drut} have simulated a graphene field theory with
staggered lattice fermions in which electrostatic degrees of freedom are formulated on a
(3+1)-dimensional lattice, while the electron fields are restricted to a 
(2+1)-dimensional slice. Their results favour the scenario that suspended
graphene with $\alpha_{\rm eff}=2.16$ is an insulator. 
More recent simulations with an improved 
fermion action support this scenario \cite{Giedt}.  Two of us 
\cite{Hands2008} have simulated an entirely 2+1-dimensional model which is in essence
a non-covariant form of the Thirring model~\cite{HandsThirring}, and showed that at infinite coupling
for $N_f < N_{fc}=4.8(2)$ graphene is an insulator, whereas for $N_f>N_{fc}$ it
is a semimetal. The results from simulations of the same model at finite
coupling provided evidence that graphene in vacuum is an insulator \cite{Armour}
in agreement with \cite{Drut, Giedt}.  More recently, the authors of
\cite{Brower} presented preliminary results from Monte Carlo simulations of the
tight-binding Hamiltonian on a hexagonal lattice. 

At nonzero temperature, universality arguments imply that the critical
properties of a $(d+1)$-dimensional theory coincide with those of a
$d$-dimensional classical spin model with the same symmetries.  The contribution
of non-zero Matsubara modes can be absorbed into non-universal aspects of the
transition. Consequently, fermions which satisfy antiperiodic boundary
conditions and do not have zero modes are expected to decouple from the scalar
sector. The validity of the dimensional reduction was confirmed with accuracy 
in Monte Carlo simulations of fermionic field theories such as the $(2+1)d$ Gross-Neveu model \cite{GNM3}
and the (3+1)d Nambu$-$Jona-Lasinio (NJL) model \cite{NJL4} and strong coupling QCD \cite{strong.c}.

There has been compelling experimental evidence \cite{Ong} that at constant low
temperature graphene undergoes a Berezinskii-Kosterlitz-Thouless (BKT) phase
transition \cite{BKT} when the intensity of an external magnetic field is
varied.  The authors of \cite{Ong} showed that the steep increase in the
electrical resistance at the Dirac point as a function of the magnetic field
fitted accurately the essential scaling relation of the BKT scenario. The BKT
transition occurs in two-dimensional systems with a U(1) symmetry and is
driven by the unbinding of vortices, as reviewed in Sec.~\ref{sec:Theory}. 
The transition separates two 
phases, neither of which have long-range order: 
a low temperature spin-wave phase where vortices and antivortices form
bound states and a high temperature plasma-like phase of unbound vortices and
antivortices.  An  analytical approach based on solutions of self-consistent
Schwinger-Dyson equations \cite{{Khveshchenko}} predicted that the critical
temperature is given by $T_{BKT}=\pi \Upsilon(T_{BKT})/2 \approx \Delta_0/8$, where
$\Upsilon(T_{BKT})$ is the helicity modulus or stiffness of the order parameter
at the transition temperature and $\Delta_0$ is the fermion mass gap at $T=0$.
However, care is needed since as shown in  \cite{Aleiner}, in a model of
graphene in which the full global symmetry is
U(4) (expected for QED$_3$ with $N_f=2$) instead of U(1), 
the creation of ``half-vortices'' is energetically more
favourable over the usual vortices. 
As a result, the critical temperature is driven to
a lower value $\tilde T_{BKT}=\pi\Upsilon(T_{BKT})/8=T_{BKT}/4$. 

In this paper we present results
from simulations of our Thirring-like graphene model~\cite{Armour} at non-zero
temperature.
As we show in Sec.~\ref{sec:formulation} on the lattice the remnant of the 
U(4)/U(2)$\otimes$U(2)  
 manifold in
which the order parameter of the continuum theory assumes values 
in U(1); we therefore do not anticipate the existence of half
vortices in our lattice model away from the continuum limit. 

The temperature in the simulation is given by $T=1/N_t a_t$, where
$N_t$ is the lattice temporal extent and $a_t$ the temporal lattice spacing. In
a model with anisotropic interactions we anticipate that the temporal ($a_t$) and
spatial ($a_s$) lattice spacings are not equal for arbitrary interaction
coupling, i.e. the anisotropy factor $a_s/a_t$ is renormalised by
quantum corrections governed by an action which treats
time and space on a different footing. This has to be taken into account
whenever deriving relations between physical quantities based on lattice
observables; fortunately for the current study all quantities can be expressed
in units of the temporal lattice spacing $a_t$.

Furthermore, as we show in Sec.~\ref{sec:TBKT} the transition temperature in natural units
is very low: i.e. $T/\Delta_0\ll1$.
This drives the critical coupling at which the BKT phase
transition occurs to a very strong value (close to the strong coupling limit)
even when the temporal lattice size $N_t=32$. This value of $N_t$ is much
larger than the values $N_t=6,\ldots,10$ usually used in simulations of nonzero
temperature QCD, and makes the study of the BKT scenario in graphene a
computationally very difficult problem.  
On the basis of large-$N_f$ arguments~\cite{Hands2008}, we believe that
at very strong couplings our Thirring-like model should become similar to the
instantaneous Coulomb interaction model~\cite{Son,Drut}.  

The
main goals of this work are: (i) to measure  $T_{BKT}/\Delta_0$; 
(ii) to obtain a first measurement of the helicity modulus $\Upsilon(T)$ for $T>T_{BKT}$
and to compare with theoretical expectations;
(iii) to measure the fermion mass gap $\Delta_T$ for $T>T_{BKT}$ and to
demonstrate that it remains nonzero even in the absence of long-range order
through exciton condensation -- this situation, which has been discussed
theoretically in the context of the Gross-Neveu model~\cite{Babaev}, is
qualitatively similar to
the pseudogap phase  observed in the phase diagram of cuprate superconductors
below optimal doping.

The paper is organised as follows:
In Sec.~\ref{sec:formulation} we present both the  continuum model and the lattice formulation
used here, along with a discussion of its global symmetries and breaking
patterns.  In Sec.~\ref{sec:Theory} we briefly review the classic BKT theory of the thermal
phase transition in planar models with U(1) global symmetry, and discuss
modifications if the global symmetry is expanded.
In Sec.~\ref{sec:results}  we present our simulation results,
and in Sec. 5 we summarise and discuss our conclusions.

\section{Formulation of the Model}
\label{sec:formulation}
Our starting point is a model of relativistic Dirac fermions moving in 2+1
dimensions and interacting via an instantaneous Coulomb interaction. In
Euclidean metric the action is~\cite{Gorbar,Son,Khveshchenko}:
\begin{equation}
S_1=\sum_{a=1}^{N_f}\int dx_0d^2x(\bar\psi_a\gamma_0\partial_0\psi_a
+v_F\bar\psi_a\vec\gamma.\vec\nabla\psi_a+iV\bar\psi_a\gamma_0\psi_a)
+{1\over{2e^2}}\int dx_0d^3x(\partial_i V)^2,
\label{eq:model}
\end{equation}
where $e$ is the electron charge, $v_F$ the Fermi velocity, $V$ the electrostatic potential, and the
$4\times4$ Dirac matrices satisfy $\{\gamma_\mu,\gamma_\nu\}=2\delta_{\mu\nu}$,
$\mu=0,\ldots,3$ (note $\gamma_3$ does not appear in (\ref{eq:model})). For monolayer graphene 
the number of fermion flavours is $N_f=2$. 

For sufficiently large coupling $e^2$ the description in terms of
massless relativistic excitations may be disrupted by condensation of bound
fermion-hole exciton pairs in the ground state, signalled by an order parameter
$\langle\bar\psi\psi\rangle\not=0$, with the result that a gap appears in the
fermion spectrum, corresponding to a transition from a conductor
to an insulator. The spontaneously broken global symmetry is U($2N_f$)
generated by rotations of the form $\psi\mapsto UV\psi$, $\bar\psi\mapsto\bar\psi
U^\dagger\gamma_3\gamma_5 V^\dagger\gamma_5\gamma_3$, with $U$ acting on flavour
indices $a=1,\ldots,N_f$ and $V$ a $2\times2$ matrix generated by the set
$\{\One,\gamma_3,\gamma_5,i\gamma_3\gamma_5\}$, where
$\{\gamma_\mu,\gamma_5\}=0\;\forall\mu$.  The order parameter remains
invariant under independent U($N_f$) rotations generated by
both $\One$ and $i\gamma_3\gamma_5$, resulting in a breaking pattern
\begin{equation}
\mbox{U}(2N_f)\to\mbox{U}(N_f)\otimes\mbox{U}(N_f).
\label{eq:contsym}
\end{equation}

At zero temperature, for $N_f<N_{fc}$
the model predicts a finite sequence of
quantum critical points (QCPs) whose properties at the critical coupling
$e^2_c(N_f)$ depend on
$N_f$~\cite{Son}. The ground state is then an excitonic condensate for
$e^2>e_c^2$. Numerical simulations of the lattice model described
below find $N_{fc}\simeq5$~\cite{Hands2008}. The QCP is an ultraviolet-stable
fixed point of the renormalisation group, implying a divergent correlation
length and algebraic behaviour of correlation functions which in principle may
be distinct from that of free-field theory. If the physical value of $e^2$
in graphene were numerically close to the fixed-point value, in either
subcritical or supercritical regimes, then the QCP might dominate the behaviour
of low energy charged excitations, with profound impact on the description of
transport.
Ultimately this must be settled by experiment.

The possible relevance of a QCP has motivated the application of lattice gauge
theory simulation techniques to the study of graphene. In this paper, we study
a model discretised on a $2+1$ dimensional Euclidean cubic lattice with action which for
$N_f=2$ can be
written in the staggered fermion formulation in 
the form (with bare Fermi velocity $v_F=1$) \cite{Hands2008,Armour}:
\begin{equation}
S_{latt}={1\over2}\sum_{x\mu i}
\bar\chi^i_x\eta_{\mu x}(1+i\delta_{\mu0}V_{
x})\chi^i_{x+\hat\mu}
-\bar\chi^i_x\eta_{\mu x}(1-i\delta_{\mu0}V_{
x-\hat0})\chi^i_{x-\hat\mu}
+m\sum_{xi}\bar\chi^i_x\chi^i_x+{1\over4g^2}\sum_{x}V_x^2.
\label{eq:Thir}
\end{equation}
Here $\chi$, $\bar\chi$ are single component Grassmann fermion fields defined on
lattice sites, $m$ an artificial mass gap introduced to regularise IR
fluctuations on a finite system volume, and $V$ a boson field, which 
mimics the electric potential of (\ref{eq:model}) in the limit 
$g^2 \to \infty$, defined on the links emanating
from the sites in the timelike direction. The Kawamoto-Smit phases
$\eta_{\mu x}=(-1)^{x_0+\cdots+x_{\mu-1}}$ are lattice analogues of the Dirac
$\gamma$-matrices.
Note that $V_x$ couples to a charge density
$J_{0x}$ which is the timelike component of a conserved
current $J_{\mu x} ={i\eta_{\mu x}\over2}[\bar\chi_x\chi_{x+\hat\mu}
+\bar\chi_{x}\chi_{x-\hat\mu}]$.
Since $V$ appears in Gaussian form it may be integrated out to yield a model
of self-interacting fermions resembling the Thirring model, with a local interaction
term of the form $g^2J_{0x}^2$.
For finite $g^2$ the $V$ field couples to a light, tightly-bound electron-hole
meson~\cite{HandsThirring}, which becomes massless in the limit
$g^2\to\infty$~\cite{Hands2008} yielding identical dynamics to the electric
potential of the gauge theory (\ref{eq:model}). The simulation results presented in
Sec.~\ref{sec:results} were obtained not far from this limit.

A distinct model, with an identical $(2+1)d$ fermion sector this time interacting with
abelian lattice gauge fields defined on a (3+1)-dimensional lattice, has been studied by
Drut and L\"ahde~\cite{Drut}. Their formulation is designed to reproduce the
action (\ref{eq:model}), which describes a long-ranged Coulomb interaction
between charges. Two comments about the relation between the models are worth
making:
\begin{itemize}

\item
The fermionic sectors share the same global symmetries. In the weakly coupled
long-wavelength
limit (\ref{eq:Thir}) describes $N_f=2$ four-component Dirac fermions~\cite{BB}. 

\item
The continuum theories modelled coincide in the strong coupling
($e^2,g^2\to\infty$)  and/or large-$N_f$ limits.

\end{itemize}
In particular, the estimate $N_{fc}=4.8(2)$ obtained using (\ref{eq:Thir}) is
expected to hold for both models~\cite{Drut, Hands2008}.

Next we discuss symmetry breaking in the model
(\ref{eq:Thir}).  In the limit
$m\to0$ there is a global ``chiral'' symmetry 
\begin{equation}
\chi_x\mapsto\exp
(i\alpha\varepsilon_x)\chi_x;\;\;\; \bar\chi_x\mapsto\exp (i\alpha\varepsilon_x)\bar\chi_x
\label{eq:U1e}
\end{equation}
where $\varepsilon_x\equiv(-1)^{x_0+x_1+x_2}$, the lattice analogue of
$\gamma_5$, 
distinguishes
odd and even sublattices.
For $N$ species of lattice fermion corresponding to $N_f=2N$ continuum flavours,
excitonic condensation of the form
$\langle\bar\chi\chi\rangle\equiv V^{-1}\partial\ln{\cal Z}/\partial m\not=0$ 
(${\cal Z}$ is the
partition function on the Euclidean spacetime lattice) induces
a spontaneous symmetry breaking of the form
\begin{equation}
\mbox{U}({N_f/2})\otimes\mbox{U}({N_f/2})\to\mbox{U}({N_f/2}). 
\end{equation}
Only in the
weak-coupling continuum limit must we necessarily expect a restoration of the
continuum breaking pattern (\ref{eq:contsym}), implying in particular 
that ${7\over4}N_f^2$ would-be Goldstone modes remain massive for non-zero
lattice spacing~\cite{Hands:2004bh}. At the QCP, however, weak
coupling cannot be assumed; moreover the effective theory need not even be
Lorentz invariant. It remains unclear, therefore, whether the enhanced symmetry 
of (\ref{eq:model})
will be fully restored, and a more systematic study of the discretised action
as advocated in \cite{Giedt} will ultimately be needed to resolve this issue.

Finally, we mention an important technical issue concerning the model
(\ref{eq:Thir}) which does not apply to the gauge-theory formulation~\cite{Drut}.
For the action (\ref{eq:Thir}) there is no symmetry guaranteeing transversity of 
the vacuum polarisation tensor (i.e. $\Delta^-_\mu\Pi_{\mu\nu x}\not=0$, where $\Delta^-_\mu$ 
is the
backward difference operator),
resulting in an additive renormalisation of the coupling $g^2$:
\begin{equation}
g^2_R={g^2\over{1-g^2/g^2_{\rm lim}}}, 
\end{equation}
where $g^2_{\rm lim}(N_f)<\infty$ defines the effective location of the strong
coupling limit. Unitarity is violated for $g^2>g^2_{\rm lim}$. 
In refs. \cite{Hands2008,Christofi2007} $g^2_{\rm lim}$ was
identified numerically with $g^{-2}_{\rm peak}$ defined by
the ($m$- and volume-independent) 
location of a peak in the order parameter
$\langle\bar\chi\chi\rangle$ found in the broken symmetry phase.

\section{Theoretical Expectations at Nonzero Temperature}
\label{sec:Theory}

In the excitonic phase which forms at $T=0$ for $g^2>g_c^2$, for $N_f=2$ the order parameter
$\langle\bar\chi\chi\rangle\equiv\phi=\phi_0e^{i\theta}$ 
spontaneously breaks a U(1) global symmetry of the
action (\ref{eq:Thir}). For $T>0$ long-range order is forbidden by the
Coleman-Mermin-Wagner theorem \cite{coleman}; rather, we expect at low $T$ a phase where
low energy phase fluctuations are described by an effective Hamiltonian
\begin{equation}
H_{eff}\propto{1\over2}(\vec\nabla\phi)^* \cdot (\vec\nabla\phi)\approx
{\Upsilon\over2}(\vec\nabla\theta)^2, 
\label{eq:Heff}
\end{equation}
where in this context the parameter $\Upsilon$ is called the {\em helicity
modulus\/}, and
correlation functions decay algebraically:
\begin{equation}
\lim_{m\to0}\langle\phi(0)\phi^\dagger(r)\rangle
=\phi_0^2\langle e^{i\theta(0)}e^{-i\theta(r)}\rangle
\propto r^{-\eta},
\end{equation}
with critical exponent $\eta=T/(2\pi\Upsilon)$.
As temperature rises topologically non-trivial excitations become important. A
vortex of charge $q$ has the form (in polar coordinates $r,\psi$) $\theta=q\psi$,
$\vert\vec\nabla\theta\vert=q/r$, and energy 
\begin{equation}
E_q=\pi\Upsilon q^2\ln{L_s\over a_s},
\label{eq:Evor}
\end{equation}
where $L_s$ is the spatial extent of the universe and $a_s$ the lattice spacing.
Overall charge neutrality is thus a requirement at low $T$ if $E$ is to remain
finite. Since a vortex can be located at any one of $(L_s/a_s)^2$ (dual) lattice
sites,
the entropy
\begin{equation}
S=2\ln{L_s\over a_s}.
\end{equation}
The free energy $F=E-TS$ of a $\vert q\vert=1$  vortex thus changes sign at a critical
temperature
\begin{equation}
T_{BKT}={\pi\over2}\Upsilon.
\label{eq:TBKT}
\end{equation}
This is the celebrated Berezinskii-Kosterlitz-Thouless transition ~\cite{BKT}
between a
low-$T$ critical phase in which vortices can only exist in tightly-bound dipole
pairs, and a gapped phase where unbound vortices form a ``topological
plasma'' which screens the long-range inter-vortex interaction. 

The relation (\ref{eq:TBKT}) remains true in a more sophisticated renormalisation
group treatment~\cite{Nelson}, except that $\Upsilon$ must be replaced by its screened value 
$\Upsilon(T_{BKT})$ exactly at the transition. The critical exponent $\eta$ 
describing correlations for $T<T_{BKT}$ thus obeys
\begin{equation}
\eta<\eta_c={1\over4}.
\end{equation}
A related exponent $\delta$ describes the response of the order
parameter to a small symmetry-breaking explicit mass gap $m$ via
$\langle\phi\rangle\propto m^{1\over\delta}$. It is related to $\eta$ via the
hyperscaling relation $\delta=(4-\eta)/\eta$, yielding
\begin{equation}
\delta>\delta_c=15.
\label{eq:delta}
\end{equation}

This picture may need modification when applied to (\ref{eq:model}).
Aleiner {\it et al\/}~\cite{Aleiner} have performed a similar analysis for the
U(2)-valued $\langle\bar\psi\psi\rangle$ using a Hamiltonian 
with independent moduli for U(1)- and SU(2)-valued  fluctuations of the
order parameter field. The crucial point is that the SU(2) sigma model is
asymptotically free, implying that $\Upsilon_{\rm SU(2)}$ rapidly runs to zero
as high-momentum modes are integrated out, with the result that the U(1)
effective Hamiltonian (\ref{eq:Heff}) is adequate for describing physics
at large distances. However, the richer symmetry of the order parameter permits
the existence of a new kind of topological excitation called a half-vortex with
$q=\pm{1\over2}$, whose energy is still given by (\ref{eq:Evor}), and which 
is thus much more readily formed by thermal fluctuations. The BKT
transition temperature is accordingly modified to
\begin{equation}
\tilde T_{BKT}={\pi\over8}\Upsilon,
\end{equation}
with new values $\eta_c={1\over16}$ and $\delta_c=63$.

\section{Numerical Results}
\label{sec:results}

In this section we present results from our numerical investigation of the model
discussed in the previous section at nonzero temperature. More specifically we
estimate the physical critical temperature, detect fermion mass generation in
the high temperature phase and study the behaviour of $\Upsilon$ at high $T$.  In
Euclidean field theory the temperature $T$ is related to the time-extent $L_t$ of
the universe via $T=L_t^{-1}=(N_t a_t)^{-1}$ where in the second step a timelike
lattice spacing $a_t$ is specified. In general numerical simulations are
performed with $N_t$ fixed, so that $T$ is varied through variation of
$a_t(g^2)$. Since $a_t\to0$ at the QCP located at the bulk critical point
$g_c^2$, we deduce that in the semimetal phase the range $0<T<\infty$ maps to
the range $0<g^2<g_c^2$, whereas in the insulating phase the same temperature
range is mapped to $\infty>g^2>g^2_c$. In this paper we are concerned with the
latter case; bearing in mind the usual convention of presenting results in terms
on inverse coupling, and also the additive coupling renormalisation described in
the previous section, we will therefore be working in the range $g^{-2}_{\rm
lim}<g^{-2}<g_c^{-2}$.

\subsection{BKT Transition}
\label{sec:TBKT}
The first set of simulations were performed with a lattice temporal extent
$N_t=16$ and spatial extents $N_s=32, 48$. For these lattice volumes $g^{-2}_{\rm
peak} \approx 0.375$; recall that the value $g^{-2}_{\rm lim}$ corresponding 
to the infinite coupling limit has previously been identified with $g^{-2}_{\rm
peak}$. 
However, this
value of $g^{-2}_{\rm peak}$ is higher than the value $g^{-2}_{\rm
peak}\approx 0.30(2)$ found at $T=0$~\cite{Hands2008}.
Although the
existence of $g^{-2}_{\rm peak}$ defining the effective strong coupling limit
is a ultraviolet (UV) artifact and therefore should not
depend on $N_t$, when $N_t$ is comparable to the lattice
spacing $a_t$, i.e. the UV scale becomes comparable to the IR scale, then 
it becomes difficult to disentangle the bulk and thermal transitions.
\begin{figure}[htb]
    \centering
    \includegraphics[width=10.0cm]{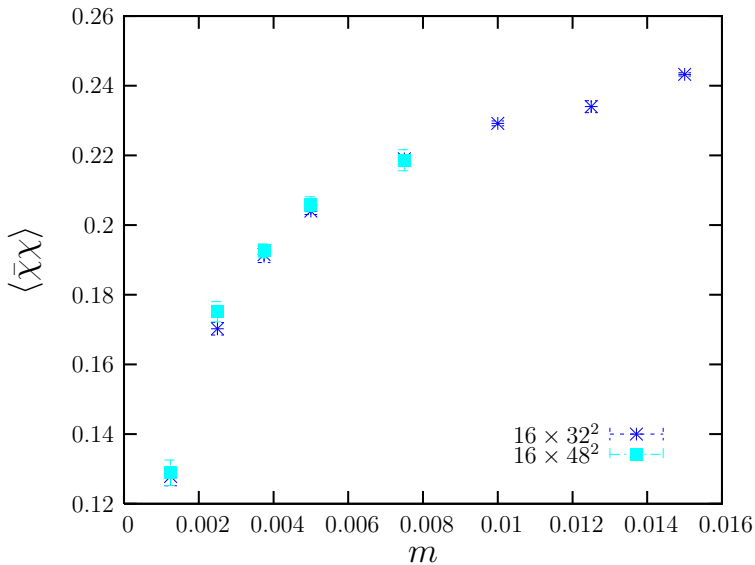}
    \caption{(color online)
Exciton condensate 
$\langle\bar\chi\chi\rangle$ versus $m$ from simulations at $g^{-2}=0.375$ 
on $16\times32^2$ and $16\times48^2$ lattices.} 
   \label{gr:fig1}
\end{figure}
In Fig.~\ref{gr:fig1} we present results for the
exciton condensate $\langle\bar\chi\chi\rangle$
versus $m$ for $N_s=32, 48$ and $g^{-2}=0.375$. It appears that
finite volume effects are negligible down to $m=0.00125$. We then fitted the
data at $g^{-2}=0.375, 0.400$ from simulations on a $16 \times 32^2$ lattice to
the scaling relation:
\begin{equation}
\langle \bar{\chi}\chi\rangle = Cm^{1/\delta}.
\label{eq:eq1}
\end{equation} 
At the critical temperature $T_{\rm BKT}$ we expect $\delta=15$. The results for the
exponent $\delta$ and the fit qualities ($\chi^2/\mbox{dof}$) are presented in
Table~\ref{tab:t1}. The data and the fitted curves are shown in
Fig.~\ref{gr:fig2}. The very low fit qualities and the values of $\delta=5.5(1),
5.1(1)$ for $g^{-2}=0.375$ and $0.400$, respectively, imply that even at
$g^{-2}_{\rm peak}$ the temperature is higher than $T_{\rm
BKT}$: we can never go down to $T_{\rm BKT}$ in simulations with
$N_t=16$. 
\begin{table}[htb]
\centering \caption{Results from fits of $\langle\bar{\chi}\chi\rangle$ vs $m$
from simulations on $16\times32^2$ lattices.}
\medskip
\label{tab:t1}
\setlength{\tabcolsep}{1.5pc}
\begin{tabular}{ccc}
\hline \hline
$g^{-2}$     &   $\delta$  &  $\chi^2/dof$      \\
\hline
0.375       &  5.5(1)  &  30     \\
0.400       &  5.1(1)     &  31     \\
\hline \hline
\end{tabular}
\end{table}
\begin{figure}[htb]
    \centering
    \vspace{10mm}
    \includegraphics[width=10.0cm]{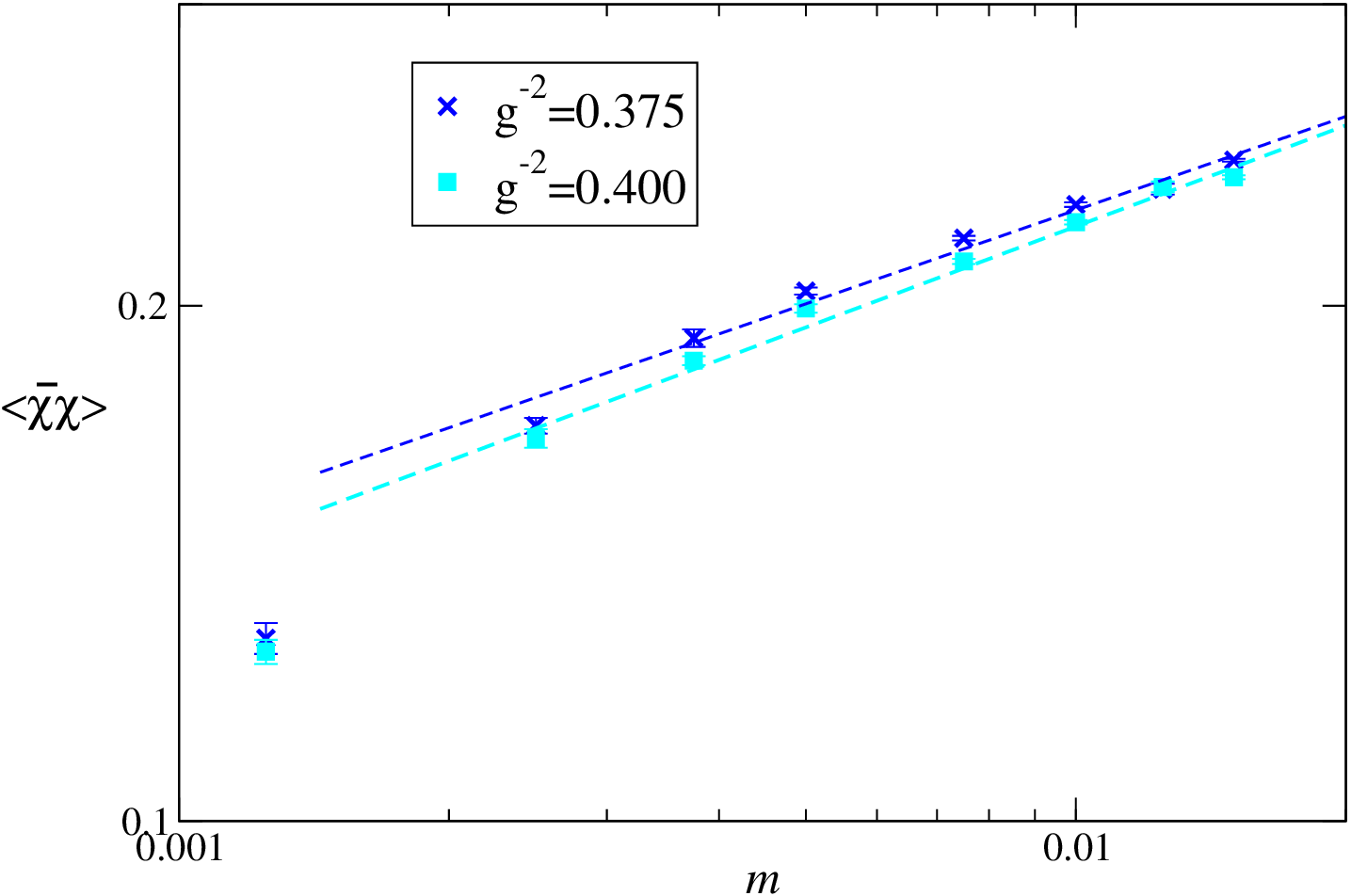}
    \caption{(color online)
$\langle\bar\chi\chi\rangle$  versus $m$ from a $16\times32^2$ lattice.}
   \label{gr:fig2}
\end{figure}
\begin{figure}[htb]
    \centering
    \vspace{10mm}
    \includegraphics[width=10.0cm]{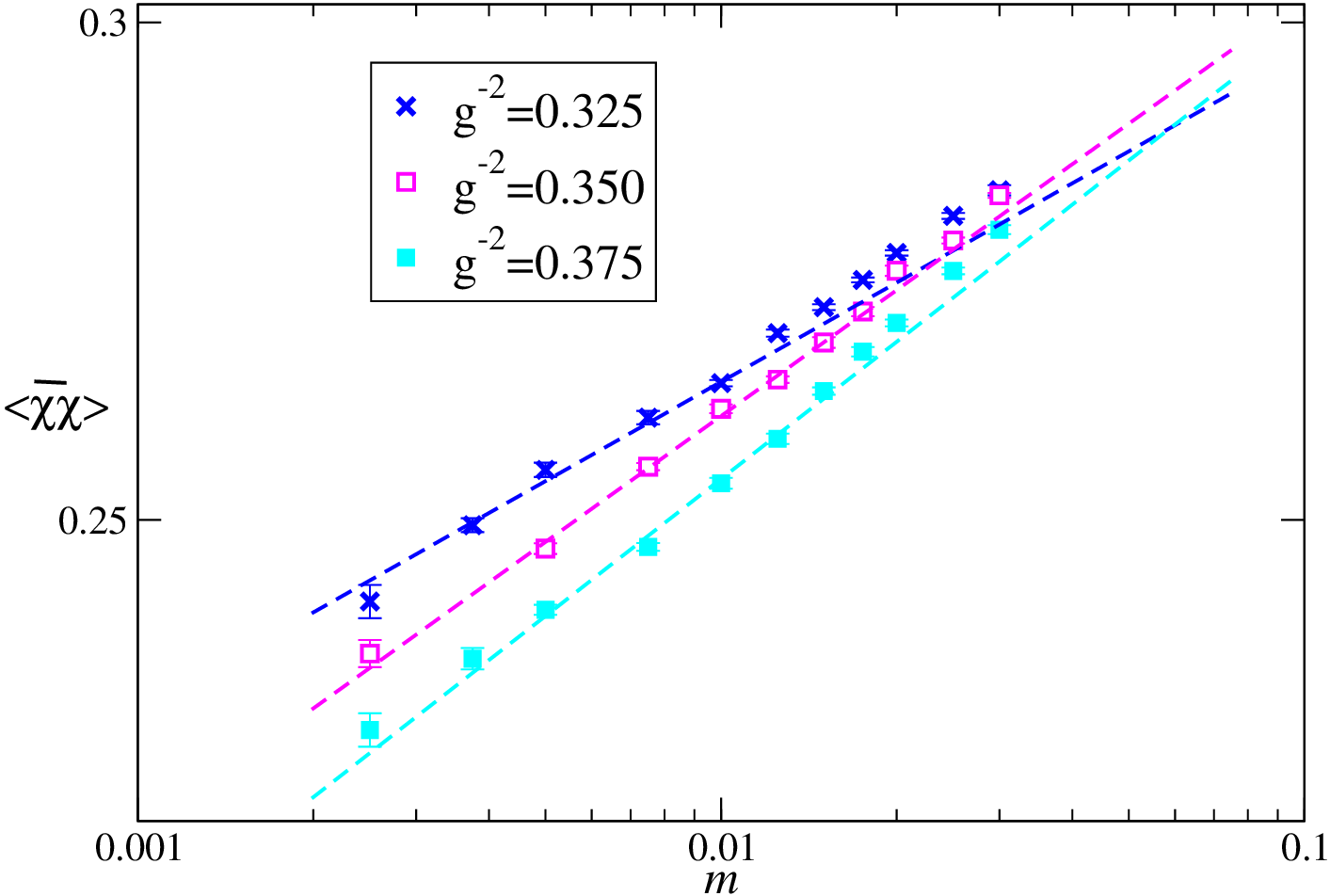}
    \caption{(color online)
$\langle\bar\chi\chi\rangle$ versus $m$ from a $32\times64^2$ lattice.}
   \label{gr:fig3}
\end{figure}

\begin{table}[htb]
\centering \caption{Results from fits of $\langle\bar{\chi}\chi\rangle$ vs $m$ from
simulations on $32\times64^2$ lattices.}
\medskip
\label{tab:t2}
\setlength{\tabcolsep}{1.5pc}
\begin{tabular}{ccc}
\hline \hline
$g^{-2}$   &   $\delta$  &  $\chi^2/dof$      \\
\hline
0.325     &   19.1(8)   &   1.7    \\
0.350     &   15.0(3)   &   1.5     \\
0.375     &   13.8(3)   &   3.9    \\
\hline \hline
\end{tabular}
\end{table}
These preliminary simulations teach us that it will require very large lattices
to identify a BKT transition.  In order to approach $T_{\rm BKT}$ we tried
$N_t=32$ and $N_s=64$. The simulations on such a large lattice at strong
couplings required enormous computational time 
because the number of iterations of the conjugate gradient algorithm
required for the inversion of the Dirac matrix kernel of (\ref{eq:Thir})
increased dramatically.  For this reason it has not proved possible to identify a
transition via singular behaviour of the susceptibility
$\partial\langle\bar\chi\chi\rangle/\partial m$ or the specific heat as was
done, say, for fermion pairing leading to long-ranged phase coherence in the
$(2+1)d$ Gross-Neveu model~\cite{Hands2001}, with
$T_{BKT}/\Delta_0\approx0.5$, using $N_t=4$, $N_s=30,\ldots,150$. 

Our strategy for locating $T_{BKT}$ is therefore based entirely on the critical
scaling relation (\ref{eq:eq1}).
The data
for $\langle \bar{\chi}\chi \rangle$ versus $m$ were fitted to
(\ref{eq:eq1}) for the ranges $m=0.0025,...,0.010$ for $g^{-2}=0.325$,
$m=0.0025,...,0.0175$ for $g^{-2}=0.350$ and $m=0.0025,...,0.015$ for
$g^{-2}=0.375$. The results are presented in Table~\ref{tab:t2} and 
Fig.~\ref{gr:fig3} shows the data and the fitted curves.  The value of
$\delta=15.0(3)$ found at $g^{-2}=0.350$ implies that the BKT transition occurs at this
coupling. It increases to $19.1(8)$ at
$g^{-2}=0.325$ which corresponds to a larger lattice spacing $a_t$ and hence lower
$T$, consistent with the BKT scenario. Note also that at the lowest
temperature ($g^{-2}=0.325$) the scaling region shrinks as
compared to higher $T$ ($g^{-2}=0.350$), because as $m$ increases the system
crosses over to the $T=0$ scaling.  The slightly increased $\chi^2/dof$ for
$g^{-2}=0.375$ provides evidence that for $g^{-2} > 0.350$ the critical scaling 
based on (\ref{eq:eq1}) is not valid because this
coupling lies in the high temperature phase.

\begin{figure}[htb]
    \centering
    \includegraphics[width=10.0cm]{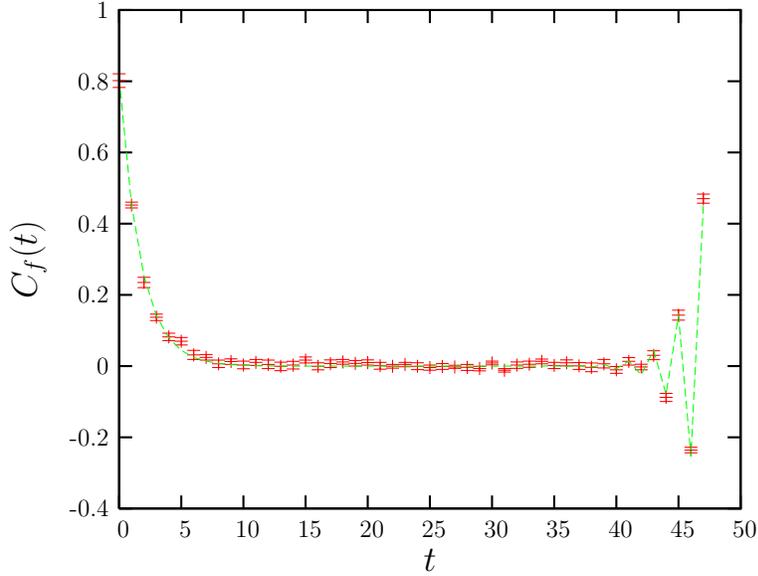}
    \caption{(color online)
Fermion correlator for $g^{-2}=0.35, m=0.01$ on a $48\times24^2$ lattice.}
   \label{gr:fig4}
\end{figure}
\begin{figure}[h]
    \centering
    \includegraphics[width=10.0cm]{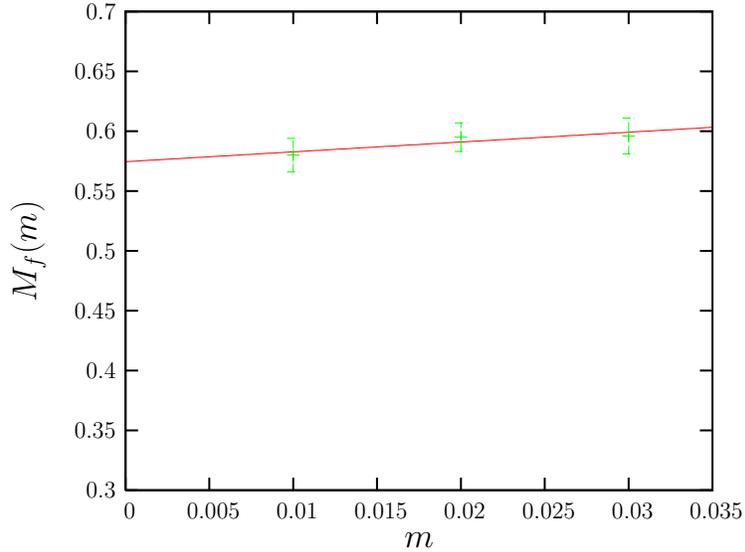}
    \caption{(color online)
Fermion mass gap  $M_f(m)$ versus $m$ from simulations with $g^{-2}=0.35,0.375$ 
on a $48\times24^2$ lattice.}
   \label{gr:fig5}
\end{figure}
In order to eliminate the lattice spacing and estimate the physical critical
temperature at the BKT transition we measured the $T=0$ fermion mass at
$g^{-2}=0.350$.  Using point sources we calculated the zero momentum 
fermion timeslice correlator 
\begin{equation}
C_f(t)=\sum_{\vec x\;{\rm even}}
\langle\chi_{\vec0,0}\bar\chi_{\vec x,t}\rangle,
\end{equation}
where ``even'' refers to sites with spatial coordinate $\vec x$ obeying
$(-1)^{x_1}=(-1)^{x_2}=1$. This restriction improves the signal to noise ratio, and
originates in the observation that the action (\ref{eq:Thir}) is invariant only
under translations by an even number of lattice spacings.
The simulations were performed on cold lattices with $N_t=48$ and
$N_s=24$ for $m=0.01, 0.02, 0.03$.  In Fig.~\ref{gr:fig4} we present the data
for $C_f(t)$ for $m=0.01$. The fermion correlator data were fitted to:
\begin{equation}
C_f(t)=A[\exp(-M_ft)-(-1)^t\exp(-M_f(N_t-t))].
\label{eq:eq2}
\end{equation}
This form assumes that the spectral density $\rho(s)$ is saturated by a pole at
$s=M_f^2$ in both particle and hole branches, appropriate for zero doping.  In
practice this assumption is justified by the quality of the fit, evident in
Fig.~\ref{gr:fig4}. The minus sign between the forward and backward terms is
due to our choice of antiperiodic boundary conditions in the timelike
direction.  The values $M_f(m)$ extracted from fits to (\ref{eq:eq2}) were
fitted to a linear scaling relation $M_f(m)=\Delta_0+a_1m$, where $\Delta_0$ is
the mass gap.  The data and the fitted line is shown in Fig.~\ref{gr:fig5}.
The extrapolation to $m=0$ at $g^{-2}=0.35$ yields $\Delta_0 a_t=0.57(2)$.  The
physical estimate for the BKT temperature is then given by: \begin{equation}
{T_{BKT}\over\Delta_0} \equiv \frac{1}{N_t \Delta_0} = 0.055(2).
\label{eq:TBKTnum} \end{equation}

This result is slightly below half of the analytical prediction 
$T_{BKT}/\Delta_0 \approx 1/8$ obtained by self-consistent solution of
Schwinger-Dyson equations in \cite{Khveshchenko}.
It is only possible to convert it into physical units indirectly, using the
estimate $\Delta_0\approx35$ meV obtained in \cite{Drut2010} by modelling the
$T$-dependence of electrical conductivity measured in suspended
graphene samples~\cite{Bolotin2008}. This yields $T_{BKT}\approx20$ Kelvin.
It should be stressed that this result has still to be extrapolated to the
continuum limit $N_t\to\infty$, $a_t\to0$. Another factor to bear in mind 
once lattice discretisation artifacts disappear is that 
the U(4) global symmetry of the continuum model (\ref{eq:model}) will be recovered.   
In that case, as described in Sec.~\ref{sec:Theory} the critical temperature  
$\tilde T_{BKT}$ will be smaller than the value (\ref{eq:TBKTnum})
by a factor of four, because half-vortices will 
become energetically favoured 
and dominate the disruption of long-range phase coherence~\cite{Aleiner}. 

\subsection{Helicity Modulus}
Next we present numerical estimates of $\Upsilon(T)$: we briefly review
the method, adapted from~\cite{Hands:2005vn}. The mass term in (\ref{eq:Thir})
is replaced by a spatially-varying source of the form 
$j\exp(i\theta(\vec x)\varepsilon_x)$, where the single-valued phase is
defined by
\begin{equation}
\theta(x_1,x_2)={{2\pi}\over N_s}(n_1x_1+n_2x_2).
\end{equation} 
The helicity modulus parametrises the response of the axial current
$J^a_{\mu x}={{i\eta_{\mu x}}\over2}[\bar\chi_x(\varepsilon\chi)_{x+\hat\mu}
+\bar\chi_x(\varepsilon\chi)_{x-\hat\mu}]$, which is conserved in the limit
$j\to0$: 
\begin{equation}
\vec J^a(j)=\Upsilon(j)\vec\nabla\theta={{2\pi\Upsilon}\over L_s}(n_1,n_2).
\label{eq:superflow}
\end{equation}
To make contact with the theoretical considerations discussed above requires the
extrapolation $j\to0$. Note that because $\vec\nabla \cdot \vec J^a$ has the same
form as the kinetic energy term in the action (\ref{eq:Thir}), the dimensionless 
variables appearing in (\ref{eq:superflow}) are $\vec J^aa_sa_t$, and $\Upsilon
a_t$, meaning that $\Upsilon$ naturally scales like a mass gap.
In practice to minimise discretisation artifacts we choose $n_1=1$, $n_2=0$.
For technical reasons associated with the 
choice $N_f=2$,  the results for $\Upsilon$
presented in this
paper were calculated in the ``partially-quenched'' approximation, in which
equilibrated field configurations were generated using a spatially-constant mass
$m$, the spatially-varying source only being 
introduced for the  measurement of $\vec J^a$.

\begin{figure}[htb]
    \centering
    \includegraphics[width=10.0cm]{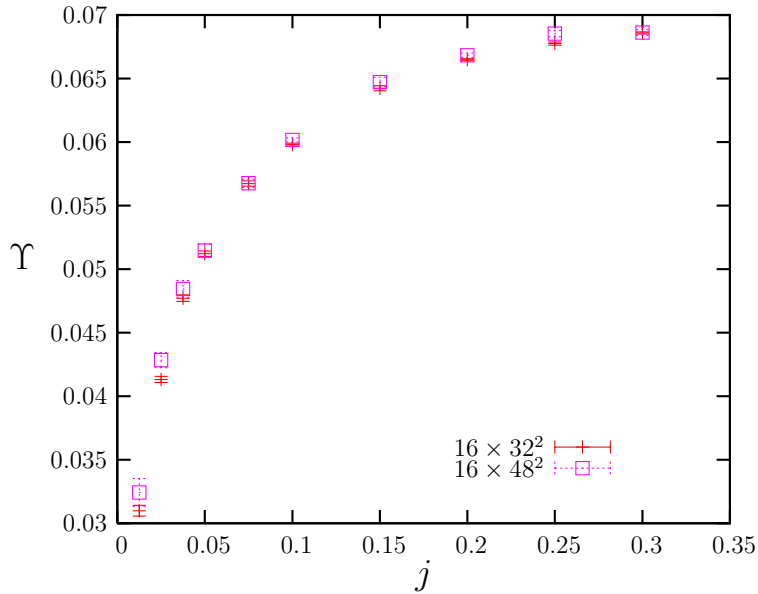}
    \caption{(color online)
$\Upsilon$ versus $j$ from simulations with $g^{-2}=0.45$, $m=0.00125$ on $16\times32^2$ 
and $16\times 48^2$ lattices.}
   \label{gr:fig6}
\end{figure}
\begin{figure}[htb]
    \centering
    \includegraphics[width=10.0cm]{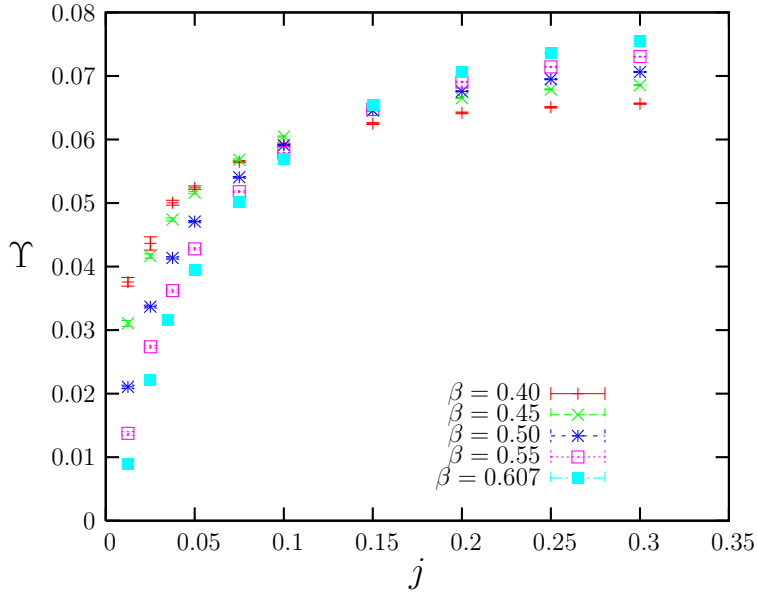}
    \caption{(color online)
Chirally extrapolated $\Upsilon$ versus $j$ for different values of $g^{-2}$ 
extracted from simulations on a 
$16\times32^2$ lattice.}
   \label{gr:fig77}
\end{figure}

Given that $\Upsilon$ is noisier than
$\langle\bar{\chi}\chi\rangle$ we restricted our simulations 
to a lattice with $N_t=16$ and were therefore only able to study high
temperatures.  In Fig.~\ref{gr:fig6} we present $\Upsilon(j)$ 
for $m=0.00125$ and $g^{-2}=0.45$ extracted from simulations with $L_s=32$
and $L_s=48$. It is inferred that effects due to finite $L_s$ are small, in
contrast to results from the Gross-Neveu model at non-zero baryon density
with $T<T_{BKT}$~\cite{Hands:2005vn}.  In order to
extract the $m=0$ value of $\Upsilon$ for each value of $j$ we
performed linear extrapolations  using $\Upsilon(m,j) =
\Upsilon(m=0,j) + a_2m$ The results for $\Upsilon(m=0,j)$ versus $j$ for
different $g^{-2}<g_c^{-2}$ corresponding to $T>T_{BKT}$ are shown in
Fig.~\ref{gr:fig77}. 

Unfortunately, we don't
have a model permitting a reliable extrapolation of these data to $j\to0$.
The data show a marked downward curvature as $j\to0$ and it is therefore
plausible, bearing in mind the insensitivity to $L_s$,
that $\Upsilon$ vanishes in this limit, as expected for $T>T_{BKT}$ 
(however the figure, including the point where curves corresponding
to differing temperatures intersect at $j\approx0.125$, is qualitatively
very similar to data taken with finite $L_s$ and fixed $T<T_{BKT}$ 
but varying baryon density in the 2+1$d$ Gross-Neveu
model~\cite{Hands:2005vn}). For $j<0.1$ there is a clear $T$-dependence. 
For reference Eqn. (\ref{eq:TBKT}) predicts 
$\Upsilon(T_{BKT})a_t=0.040$, of the same order of magnitude as $\Upsilon(j)$
around the ``knee'' seen in the data of Fig.~\ref{gr:fig11} at 
$j\sim0.1$; even though a quantitative description is still
lacking, therefore, the signal for $\Upsilon$ is broadly consistent 
with the BKT scenario outlined in Sec.~\ref{sec:Theory}.

\subsection{Quasiparticle Thermal Mass and Dispersion Relation}
\label{sec:pseudogap}
Next, we calculated the fermion thermal mass in the high temperature region from
simulations on $16 \times 32^2$ lattices.  Once again, the fact that the fermion
correlator has a smaller signal-to-noise ratio than the order parameter
$\langle\bar\chi\chi\rangle$ forces us to work on smaller volumes.
Now, at $T>0$ fermions can acquire a non-zero thermal mass even in the
absence of spontaneous symmetry breaking. 
For a weakly-coupled theory, this is simply the Debye screening mass 
$m_D\sim gT$, but in a strongly-coupled theory where dynamical mass generation
at $T=0$ results from spontaneous symmetry breaking, it is better to draw 
analogies with the ``pseudogap'' phase thought to form in
cuprate superconductors at strong coupling or low carrier density \cite{Babaev}.
Once again, we write the pairing field as $\bar\chi\chi=\phi_0e^{i\theta}$.
For a temperature range $T_{BKT}<T<T^*$,  the pseudogap phase arises
due to the  ``local'' gap modulus $\phi_0$, neutral under $U(1)$
rotations, remaining nonzero, while the phase $\theta$ fluctuates violently,
precluding both a non-zero order parameter and also the long-ranged phase coherence
signalled by a non-vanishing helicity modulus. In Ref.~\cite{Babaev} the
temperature $T^*$ in the $(2+1)d$ Gross-Neveu model is predicted to coincide
with the estimate $\Delta_0/2\ln2$ given by mean field theory, and the
difference $T^*-T_{BKT}\simeq(N_f\ln2)^{-1}$.
The
existence of the pseudogap phase at non-zero temperature was demonstrated in
numerical simulations of Gross-Neveu models with $U(1)$ \cite{Hands2001}
and $SU(2)\times SU(2)$ \cite{Strouthos} chiral symmetries, and analytically in
the $4d$ NJL model \cite{Castorina}. 

\begin{figure}[htb]
    \centering
    \includegraphics[width=10.0cm]{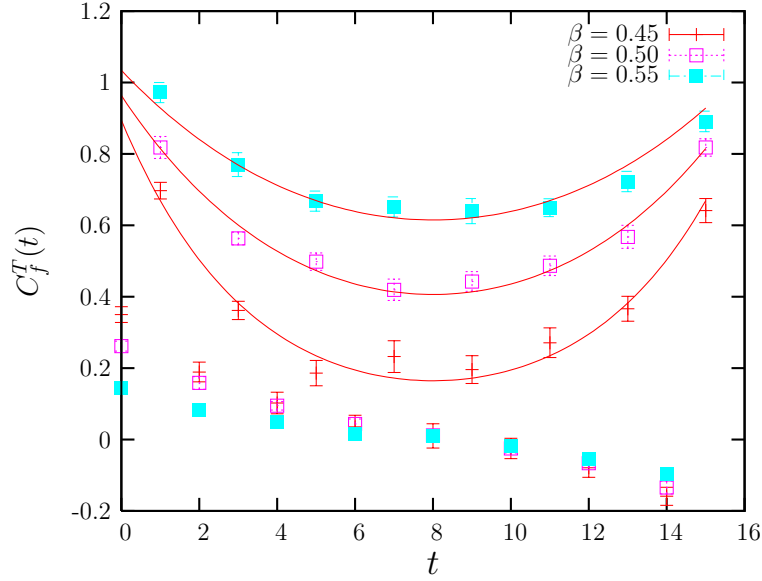}
    \caption{(color online)
Fermion correlator with $m=0.00125$ and $g^{-2}=0.45, 0.50, 0.55$ 
on a $16\times32^2$ lattice. The curves result from fits to data with $t$ odd.}
   \label{gr:fig7}
\end{figure}
In Fig.~\ref{gr:fig7} we show the 
fermion timeslice correlator $C_f^T(t)$ 
for $g^{-2}=0.45, 0.50, 55$ and $m=0.00125$.  We fitted the
data for odd timeslices only to 
\begin{equation} 
C_f^T(t)=A[\exp(-M_f^T t) +
\exp(-(N_t-t)M_f^T t)] 
\end{equation} 
The small values of $C_f^T(t)$ observed on even
timeslices signals a manifest chiral symmetry which is broken only explicitly by the
fermion bare mass term. The $U(1)_{\epsilon}$ symmetry (\ref{eq:U1e})
of staggered fermions
implies that the only nonvanishing elements of the propagator are $C_{feo}$ and
$C_{foe}$, where the $e/o$ subscripts denote sites with $\varepsilon_x=\pm1$. 

\begin{figure}[htb]
    \centering
    \includegraphics[width=10.0cm]{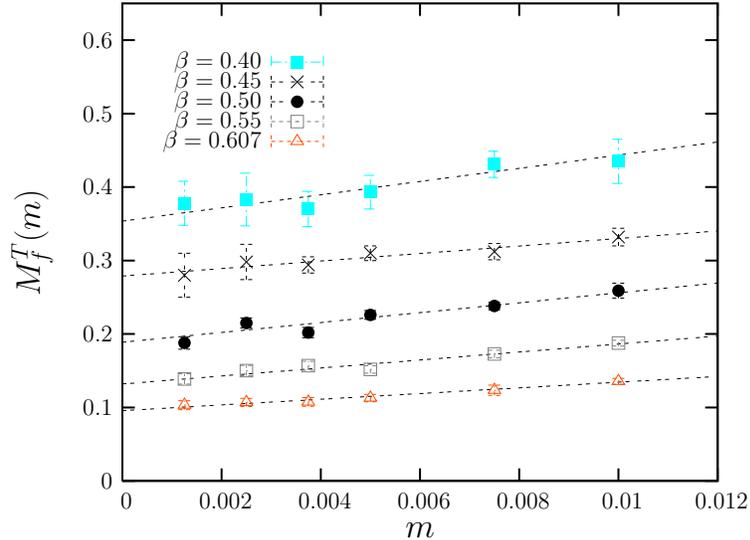}
    \caption{(color online)
Fermion thermal mass $M_f^T$ versus $m$ for various $g^{-2}$ 
extracted from simulations on a
$16\times32^2$ lattice.}
   \label{gr:fig8}
\end{figure}
\begin{figure}[htb]
    \centering
    \includegraphics[width=10.0cm]{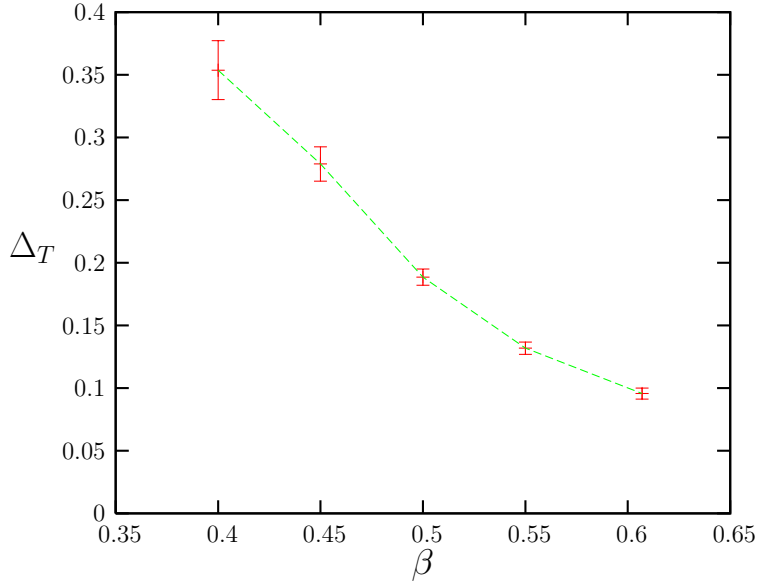}
    \caption{(color online)
Chirally extrapolated thermal mass $\Delta_T$ versus $g^{-2}$ extracted from simulations on a
$16\times32^2$ lattice.}
   \label{gr:fig9}
\end{figure}
In Fig.~\ref{gr:fig8} we present the results for $M_f^T$ versus $m$
extrapolated with a linear function $M_f^T(m)= \Delta_T+a_3m$ to the chiral limit.
Fig.~\ref{gr:fig9} shows $\Delta_T$ versus $g^{-2}$. As
$g^{-2}$ increases the lattice spacing decreases and at the bulk critical
coupling $g^{-2}_c=0.609(2)$ $a_t=a_s=0$ \cite{Armour}, implying $T \to \infty$.
It is clear from Fig.~\ref{gr:fig9} that $\Delta_T$ remains of the same order
of magnitude as $\Delta_0$ for a significant extent of
the high temperature phase $T>T_{BKT}$, lending strong support to
the pseudogap scenario with $T^*>T_{BKT}$. 

\begin{figure}[htb]
    \centering
    \includegraphics[width=10.0cm]{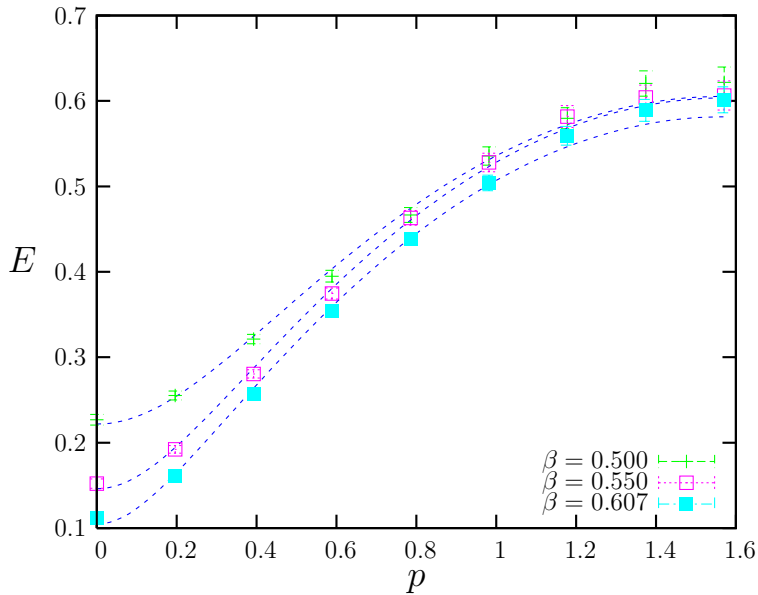}
    \caption{(color online)
Quasiparticle dispersion relation $E(p)$ as measured on a $16\times32^2$ lattice 
with $m=0.005$.}
   \label{gr:fig10}
\end{figure}

\begin{figure}[htb]
    \centering
    \includegraphics[width=10.0cm]{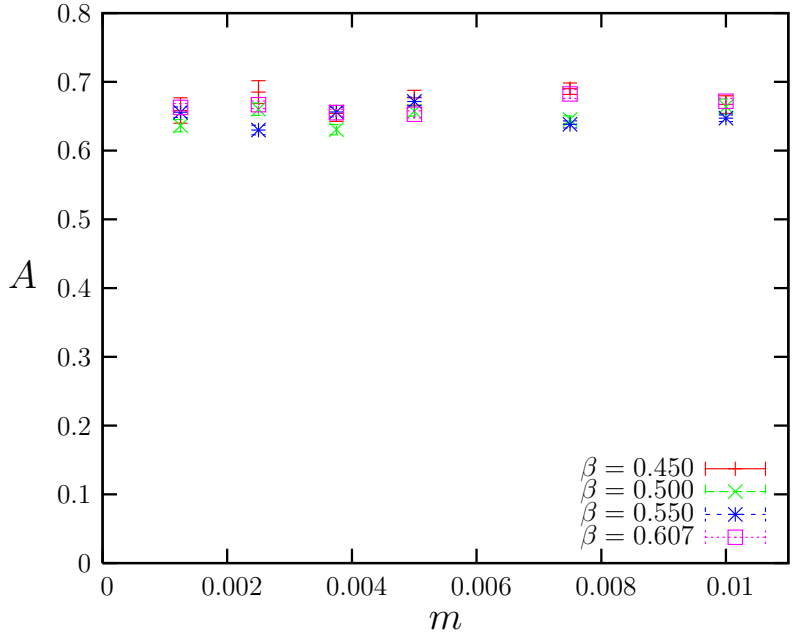}
    \caption{(color online)
The fitted parameter $A$ vs $m$ for various values of $g^{-2}$.}
   \label{gr:fig11}
\end{figure}

The fermion energy as a function of momentum is accessed via analysis of the
Euclidean timeslice propagator $C_f(\vec p,t)$ defined by
\begin{equation}
C_f^T(\vec p,t)=
\sum_{\vec x\;{\rm even}}\langle\chi(\vec0,0)\bar\chi(\vec x,t)\rangle
e^{-i\vec p.\vec x},
\end{equation}
where the components of momentum $\vec p$ take values $2\pi
n/L_s$, with $n=0,1,\ldots,L_s/4$. 
The energy $E(\vec p)$ is then
extracted by a fit of the form
\begin{equation}
C_f(\vec p,t)=B(e^{-Et}+e^{-E(L_t-t)}),
\label{eq:Efit}
\end{equation}
where again only data with $t$ odd were used.
We measured $E(\vec p)$ for $\vec p=(p_1,0)$ on   $16\times32^2$ in the high temperature
phase. 
To proceed we parametrise the dispersion relation
using
\begin{equation}
E(p)=A\sinh^{-1}(\sqrt{\sin^2p+M^2}),
\label{eq:dispfit}
\end{equation}
where for $A=1$ and $M=m$ the exact result for non-interacting
lattice fermions is recovered.
Sample fits to (\ref{eq:dispfit}) at $m=0.005$ are shown in Fig.~\ref{gr:fig10}. 
The dispersion flattens out to have zero slope at the
effective Brillouin zone edge at $p={\pi\over2}$; this flattening is a
discretisation artifact with no physical significance.
For small $M$ we can interpret $E(0)\equiv M_f^T \approx AM$ as the quasiparticle mass
(or gap), and for small $p$
in the limit $M\to0$ then $dE/dp\approx A$ is the renormalized Fermi velocity
$v_{FR}^Ta_t/a_s$ at nonzero temperature, where we have restored explicit
factors of
lattice spacing. Without further information we are unable to distinguish
between renormalization of the physical Fermi velocity and that of the cutoff
anisotropy due to quantum corrections (this point was not realised in
\cite{Armour}), but note that the latter must be $T$-independent.
Results for $A$ as a function of $m$ are shown
in Fig.~\ref{gr:fig11}.  Despite some noise in the data the parameter $A$, and
hence $v_{FR}^T$, is both $m$- and
$g^{-2}$-independent taking a numerical value $\approx 0.65$, which is very
close to the value $A \approx 0.7$ reported in \cite{Armour} at $T=0$. This
implies that the principal physical effect of the hot medium is to generate a
nonzero thermal mass, rather than to renormalize the Fermi velocity.  A similar
effect was observed in nonzero $T$ simulations of the $(2+1)d$ Gross-Neveu
model with an $SU(2) \otimes SU(2)$ chiral symmetry \cite{Strouthos}. 

\section{Summary and Conclusion}

The main result of this first, exploratory study of thermal effects in the
insulating phase of the graphene effective
theory (\ref{eq:model}) with $N_f=2$, via numerical simulation of its discrete avatar
(\ref{eq:Thir}), is the determination of the critical temperature for vortex
unbinding $T_{BKT}/\Delta_0\approx0.06$. This value is considerably smaller than
the ratio found in the Gross-Neveu model ($T_{BKT}/\Delta_0\approx0.5$)~\cite{Hands2001}, underlining the
point that different four-point Fermi interactions yield distinct dynamics in
$(2+1)d$, and that perturbative approaches such as the $1/N_f$ expansion are
unlikely to be accurate for graphene~\cite{HandsThirring}. 
It also implies that study of the BKT transition in this system is a
numerically challenging problem, requiring large lattice volumes in order to resolve the
large separation of scales. With the resources at our disposal we have been
able to work with $N_t=32$, which has enabled an estimate of $T_{BKT}$
via the
critical scaling (\ref{eq:delta}) of the order parameter with external mass source and
identification of the exponent $\delta$, but not yet, it must be stressed, via
direct observation of singular behaviour in any thermodynamic observable.
That said, it is noteworthy that our value (\ref{eq:TBKTnum}) is not too far
removed from predictions made using Schwinger-Dyson
equations~\cite{Khveshchenko}.

Two major caveats must be noted. First, predictions made using the discrete model (\ref{eq:Thir}) can
strictly only be applicable in the continuum limit; we therefore need to explore
the limit $g^2\searrow g_c^2$ to control the inevitable discretisation
artifacts, which may  scale with non-trivial powers of $a_s,\,a_t$ 
as the QCP is approached. Unfortunately in practical terms this requires the
limit $N_t\to\infty$. Secondly, as noted earlier, it is argued that in the
continuum limit the global symmetry of the effective graphene  Lagrangian
enlarges from U(1)$\otimes$U(1) to U(4), implying the existence of half-vortex
topological excitations, which exhibit an unbinding transition at a still lower
temperature $\tilde T_{BKT}=T_{BKT}/4$~\cite{Aleiner}. Since our estimate of the
critical temperature assumes the orthodox BKT scenario, we are unable to comment
further on this possibility. Resolving this question will probably require a more
refined lattice fermion discretisation, as advocated in \cite{Giedt}.

We have also presented results for the helicity modulus $\Upsilon$ as a function
of the source strength $j$ introduced to induce a circulating supercurrent in
our system. The numerical challenge has so far restricted our study
to the region $T>T_{BKT}$, but the magnitude of $\Upsilon(j)$ observed is
consistent with the expectations of the conventional BKT scenario. 
We are
unaware of any effective model enabling a controlled $j\to0$
extrapolation on finite systems.

Finally, the calculation of the quasiparticle propagator presented in
Sec.~\ref{sec:pseudogap} reveals the persistence of a gap
$\Delta_T\lapprox\Delta_0$ for temperatures $T>T_{BKT}$, despite the fact
that the form of the correlators shown in Fig.~\ref{gr:fig7} is characteristic
of propagation through a chirally-symmetric medium. As argued in
\cite{Hands2001}, in this phase the fermion flips chirality, permitting
propagation at speeds $v<v_F$, by constantly exchanging massless bosonic quanta
with the medium: this is signalled by the spectral density function $\rho(s)$ being modified from a
simple pole on the mass shell to a branch cut above the threshold at $s=\Delta_T^2$.
The situation qualitatively resembles the discussion of the pseudogap phase in
cuprates given in \cite{Babaev}. In addition, the analysis of the
fermion dispersion relation for $T>T_{BKT}$ showed that the main effect of the hot medium is 
to generate a non-zero thermal quasiparticle mass rather than to renormalize the $T=0$ physical 
Fermi velocity.

\section*{Acknowledgements}
The authors wish to thank the Diamond Light Source for kindly allowing
them to use extensive computing resources.

\end{document}